\documentclass[sigconf]{acmart}

\usepackage{listings}
\usepackage{balance}
\usepackage{fancyvrb}

\AtBeginDocument{%
  \providecommand\BibTeX{{%
    \normalfont B\kern-0.5em{\scshape i\kern-0.25em b}\kern-0.8em\TeX}}}

\copyrightyear{2021}
\acmYear{2021}
\setcopyright{acmlicensed}\acmConference[CHI '21]{CHI Conference on Human Factors in Computing Systems}{May 8--13, 2021}{Yokohama, Japan}
\acmBooktitle{CHI Conference on Human Factors in Computing Systems (CHI '21), May 8--13, 2021, Yokohama, Japan}
\acmPrice{15.00}
\acmDOI{10.1145/3411764.3445257}
\acmISBN{978-1-4503-8096-6/21/05}

\begin{document}

\setlength\leftmargini{1.5em}

\title{The Role of Working Memory in Program Tracing}

\author{Will Crichton}
\email{wcrichto@cs.stanford.edu}
\affiliation{%
  \institution{Stanford University}
}

\author{Maneesh Agrawala}
\affiliation{%
  \institution{Stanford University}
}

\author{Pat Hanrahan}
\affiliation{%
  \institution{Stanford University}
}

\renewcommand{\shortauthors}{Crichton et al.}


%
\begin{abstract}
Program tracing, or mentally simulating a program on concrete inputs, is an important part of general program comprehension. Programs involve many kinds of virtual state that must be held in memory, such as variable/value pairs and a call stack. In this work, we examine the influence of short-term working memory (WM) on a person's ability to remember program state during tracing. We first confirm that previous findings in cognitive psychology transfer to the programming domain: people can keep about 7 variable/value pairs in WM, and people will accidentally swap associations between variables due to WM load. We use a restricted focus viewing interface to further analyze the strategies people use to trace through programs, and the relationship of tracing strategy to WM. Given a straight-line program, we find half of our participants traced a program from the top-down line-by-line (linearly), and the other half start at the bottom and trace upward based on data dependencies (on-demand). Participants with an on-demand strategy made more WM errors while tracing straight-line code than with a linear strategy, but the two strategies contained an equal number of WM errors when tracing code with functions. We conclude with the implications of these findings for the design of programming tools: first, programs should be analyzed to identify and refactor human-memory-intensive sections of code. Second, programming environments should interactively visualize variable metadata to reduce WM load in accordance with a person's tracing strategy. Third, tools for program comprehension should enable externalizing program state while tracing.
\end{abstract}

\ccsdesc[500]{Human-centered computing~Laboratory experiments}

\keywords{Program tracing, program comprehension, working memory, restricted focus viewing}


\maketitle

\section{Introduction}

\begin{figure*}[t]
\centering
\begin{minipage}{0.55\textwidth}
    \centering
    \begin{lstlisting}[language=Python]
def add(n1, n2):
  output, carry = [], 0
  for d1, d2 in reversed(list(zip(n1, n2))): 
    intermediate = d1 + d2 + carry
    output.insert(0, intermediate % 10)  
    carry = 1 if intermediate >= 10 else 0
  if carry > 0:
    output.insert(0, carry)
  return output  
  
assert add([4, 7], [1, 8]) == [6, 5]\end{lstlisting}
\end{minipage}%
\begin{minipage}{0.45\textwidth}
    \centering
    \begin{lstlisting}[language=Python]
def square_twodigit(n):
  cn = min(n % 10, 10 - (n % 10))
  nmt = n + cn
  otn = n - cn
  p1 = nmt * (otn // 10 * 10)
  p2 = nmt * (otn % 10)
  sm = p1 + p2
  cn2 = cn * cn
  return sm + cn2
  
assert square_twodigit(23) == 529\end{lstlisting}
\end{minipage}
\caption{Mental procedures for addition (left) and squaring a two-digit number (right). Procedures are represented as Python programs to emphasize the similarity of mental arithmetic to program tracing.}
\label{fig:procedures}
\Description{Two Python programs are shown representing the mental computations in the cited work. The left is for mental addition, and the right is for squaring a two-digit number.}
\end{figure*}

Program tracing is the task of mentally simulating a program on concrete inputs to produce a concrete output. To trace a program, a person follows a procedure step-by-step as specified in the source code. Go here, add this, store that, repeat --- a literal human information processor. Program tracing encompasses questions such as: for some function $f$ what is the concrete value of $f(1)$? By contrast, program \textit{comprehension} involves tasks about abstract relationships, e.g.\ what is the symbolic value of $f(x)$ in terms of any $x \in \mathbb{N}$?

Prior work on human-centered design of programming tools has focused on measuring and improving comprehension, not tracing. After all, the entire point of a computer is to \textit{automatically} trace programs, and have the human operate at a higher level of abstraction. However, cognitive psychologists have repeatedly demonstrated the close relationship between action and understanding in comprehension of language\,\cite{pickering2007people} and gestures\,\cite{wilson2005case}.  By extrapolation, program tracing, the ability to \textit{execute} a computational process, is likely a component skill of program comprehension, the ability to \textit{understand} that same process. This claim has support from research in computing education and software engineering. Students who could correctly trace programs performed better on program comprehension questions\,\cite{lopez_relationships_2008,lister_further_2009}. Students' comprehension errors can often be attributed to a flawed mental model  (\textit{notional machine}) of how a computer would trace individual instructions\,\cite{sorva2007notional}. Expert programmers use tracing to understand code when its structure is not similar to a previously observed pattern (\textit{schema})\,\cite{letovsky_cognitive_1987,detienne_empirically-derived_1990}. 

Hence, to design for program comprehension, we ought to better understand what makes program tracing difficult. Consider the myriad differences between a computer tracing a program versus a person doing the same. A computer can instantaneously parse a program, effortlessly keep its runtime state in memory, and infallibly apply the language’s rules to compute the program’s output. Humans, by contrast, are much worse at these tasks. To understand why, take a moment to trace this Python program to determine what value it prints.

\begin{lstlisting}[language=Python]
def f(x, r, q):
    return r - q + x
q = 1 + 4
e = 8 - q
print(f(q, e, f(3, 5, e)))
\end{lstlisting}

While tracing this program, you may have found yourself thinking --- ``what was the value of \verb|e|? Which was the second argument in the function call? Which \verb|q| is this referring to?'' These questions all point to failures of memory to maintain program state during tracing. Specifically, failures of \textit{working memory} (WM): the ``brain system that provides temporary storage and manipulation of information''\,\cite{baddeley_working_1992}. The universal limitations of WM have been studied since the early days of cognitive psychology\,\cite{miller1956magical}: WM has a capacity of roughly $4 \pm 1$ chunks\,\cite{cowan_magical_2001}, and chunks are forgotten from WM as a function of time and interference\,\cite{oberauer_benchmarks_2018}. Mental tasks which exceed these limitations (high \textit{cognitive load}) become difficult to correctly and efficiently execute. 
Prior qualitative studies have shown that students encounter WM difficulties while tracing: they remember ``at most one value for all the variables in the program''\,\cite{vainio_factors_2007}. WM has also inspired the development of program comprehension tools such as the Code Bubbles IDE\,\cite{bragdon_code_2010}. However, to date, no work has systematically explored the role of WM in program tracing. 

As demonstrated in the example above, maintaining program state during tracing likely imposes demands on WM. Program state requires WM because it is \textit{intangible}. If a person is tracing a procedure with physical state like a recipe or a furniture manual, they can quickly survey ingredients on the kitchen counter, or find the missing parts of a half-built chair. But programs have purely virtual state --- variables, instruction pointers, call stacks --- meaning a person cannot rely on their environment to maintain state for them. Therefore in this paper, we focus on two high-level questions:

\begin{enumerate}
    \item How much program state can a person hold in WM?
    \item How do WM limitations for program state influence tracing strategies?
\end{enumerate}

WM theory suggests a few coarse answers: people can't remember much state, and they'll forget relevant state while tracing. But in this paper, our goal is to dive deeper into the nuances of each question: how much program state can a person remember in pure recall vs. interleaved with mental arithmetic? Do people linearly trace in the same fashion as a computer, or do they optimize their strategies for WM? To that end, we ran four controlled experiments to elucidate the influence of working memory for program state on tracing. In each experiment, we carefully restrict the participants' view on the program, for example by seeing a single line/function at a time, or blurring lines of code until hovered. Then we use the participants' responses and their mouse movements to derive quantitative measures of working memory influence. Our main contributions are: 





\begin{itemize}
    \item Verifying whether results from prior cognitive psychology research on similar tasks (e.g. WM capacity for paired-associates) will transfer to program tracing.
    \item Designing novel experimental methodologies to analyze participants' behavior at a fine-grained level while tracing programs.
    \item Applying WM theory to predict participants' tracing behavior, and to generate design principles for programming tools that can reduce WM load. 
\end{itemize}


\section{Background and Related Work}

To relate program tracing back to prior research on working memory, we can break down the general task of tracing into components that look more like tasks in prior work. Specifically, we will focus on how program state in tracing arises from a few basic programming concepts: arithmetic, variables, computation order, and functions.

\subsection{Arithmetic}

Mental arithmetic is a well-studied task within the context of working memory\,\cite{destefano_role_2004}. It is also a ``program'' that people trace every day --- adding multi-digit numbers is a procedure many people are taught to memorize and mentally simulate from a young age.
For example, computing $47 + 18$ is similar to tracing the program in Figure~\ref{fig:procedures} (left).

Graham Hitch\,\cite{hitch_role_1978} first found in 1978 that a working memory analysis of mental arithmetic could explain many observed calculation errors. While calculating, a person must maintain a number of \textit{intermediates} in working memory, such as all previously computed digits and a carry bit. Based on a mathematical model fit to experimental data, Hitch's theory was that a person's probability of forgetting an intermediate increased exponentially with the number of calculations since computing the intermediate. For example, in mentally computing $138 + 326$, if a person first computes $6 + 8 = 14$, then they will forget the ones-place digit $4$ with exponentially increasing probability after computing each subsequent digit.

Within the domain of mental computation, prior cognitive psychology research has focused on \textit{memorized} procedures like addition, as opposed to tracing unfamiliar procedures presented through an external medium. Zhang and Norman\,\cite{zhang1995representational} use a representational analysis to justify why Arabic numerals are an ideal number representation for performing mental multiplication. Campbell and Charness\,\cite{campbell_age-related_1991} ran an experiment where participants memorized an eight-step procedure for squaring two-digit numbers, shown in  Figure~\ref{fig:procedures} (right). They found that the majority of errors occurred within 5 stages of the procedure, and most errors could be explained as working memory errors (not calculational errors) where one intermediate was substituted for another. 


\subsection{Variables}

To understand the influence of WM on variables in program tracing, we can extrapolate the concept of an intermediate in mental arithmetic. For example, consider tracing this program:

\begin{lstlisting}[language=Python]
x = 12 + 9 - 2
print(x + 3)
\end{lstlisting}

This trace involves three kinds of intermediates. First, computing $12 + 9$ involves the intermediates of multi-digit addition, e.g. a carry bit. Second, the expression $12 + 9 - 2$ requires remembering the intermediate $21$ while computing $21 - 2$. Finally, the variable assignment $x = 19$ must be remembered while tracing the second line. The key idea is that each intermediate must be stored in WM by association to its role in the program. That association could be behavioral (carry bit), positional (left hand side of an expression), or symbolic (the letter \verb|x|). In this view, variables in program tracing are intermediates with association to strings. A person remembers \verb|x = 19|, then retrieves the value \verb|19| when cued with the variable \verb|x|.

In the vocabulary of cognitive science, remembering a variable/value pair is \textit{paired-associate learning}, and retrieving the value from memory given a variable is \textit{cued recall}. Cued recall for paired-associate learning has been studied in a variety of domains. The simplest form of study is \textit{memory span}: how many pairs can a person remember at once with no distractions? Multiple experiments on cued recall for pairs of common English nouns found that participants could remember on average 5 out of 10 pairs (when presented for 2s/pair)\,\cite{nobel2001retrieval,unsworth2009examining}.

Program tracing is not a pure memorization task, however --- a person must interleave the storage of variables in WM with the calculation of expressions. This raises the question: would calculation interfere with WM for variables? While this question has not been studied directly, its inverse has prior work: WM used by mental computation can be interfered with by certain competing tasks. In one experiment, participants had to mentally add three-digit numbers while concurrently performing a dual task. Participants were significantly more likely to forget the carry bit under a dual task which involves central executive WM (Trails task, speaking aloud an alternating sequence e.g. ``A-1-B-2-$\ldots$'') versus a dual task which involves phonological WM (articulatory suppression, repeatedly saying the word ``the'' aloud)\,\cite{furst2000separate}. 
\subsection{Computation order}

An important distinction between program tracing and mental arithmetic is that program tracing is accompanied by the program source code, while mental arithmetic uses a memorized procedure. The source code acts an external memory of the process's instructions, but also potentially for program state. For example, try tracing this program:

\begin{minipage}{\linewidth}
\begin{lstlisting}[language=Python]
x = 8
q = 3 + x
r = 6
print(x - q + r)
\end{lstlisting}
\end{minipage}

A ``linear'' strategy, the most similar to how a computer actually executes a program, would trace from line 1 to line 4, committing each variable/value pair to memory along the way. But you might have used an ``on-demand'' strategy: skim to line 4, then look up the value of each variable as needed. Either strategy produces the correct answer, although on-demand strategies nominally become more challenging in the presence of side effects (e.g.\ writing to a file) because out-of-order tracing may not produce the same ordering of effects as linear tracing.

Vainio and Sajaniemi\,\cite{vainio_factors_2007} observed that students tracing programs would adopt a strategy they called ``single value tracing.'' In our terminology, students would trace on-demand with respect to variables assigned to constants. In the example above, a student might ignore line 1 (a ``trivial'' value in their words) and skip to line 2, looking up the value of \verb|x| on demand. However, we do not know generally how often people will adopt a particular strategy, or how strategy relates to WM.

For more complex programs (e.g. with control flow, indentation, methods, etc.)  eye-tracking studies of programmers have shown that programmers will read programs non-linearly. Busjahn et al.\,\cite{busjahn2015eye} found that the gaze path of expert programmers was more similar to following control flow from top-to-bottom than to reading line-by-line like a book. A replication of Busjahn et al. by Peitek et al.\,\cite{peitek2020drives} further found that more linear programs (e.g. with fewer function calls) correlated to less vertical eye movement. Unlike these studies, we use much simpler programs and combine eye tracking with dataflow to identify WM errors from the gaze path.




\subsection{Control flow}

\begin{figure*}[t]
    \centering
    \includegraphics[width=0.7\textwidth]{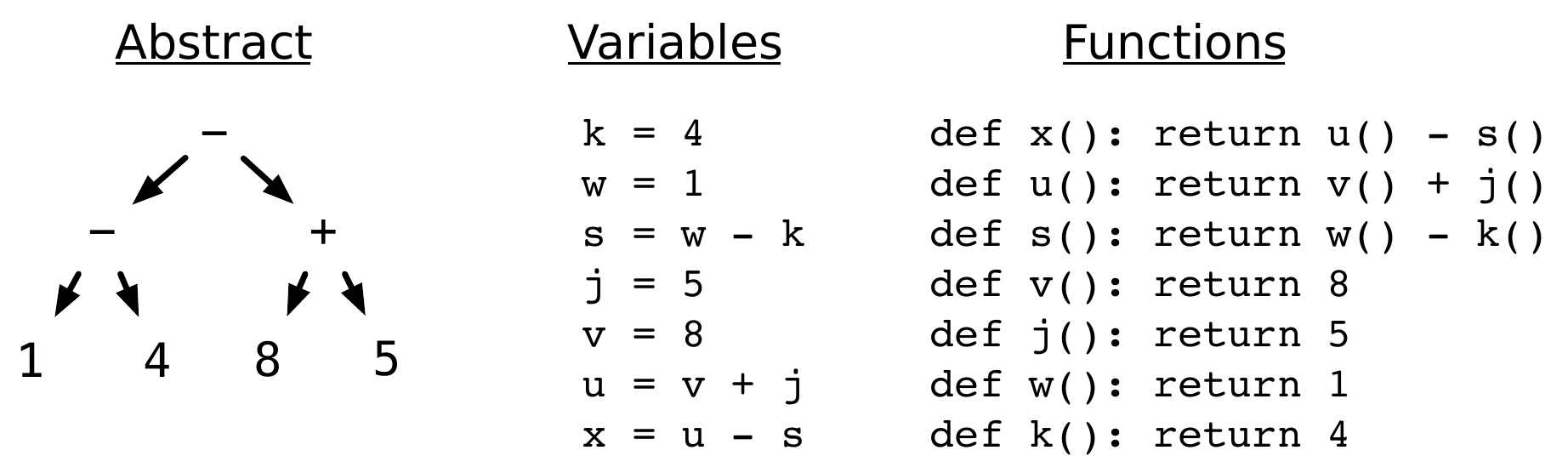}
    \caption{Three equivalent representations of the arithmetic expression $(1 - 4) - (8 + 5)$. Left is an nameless/tree representation. Center is a topological sort of the tree into a program using variables. Right is an arbitrary ordering of the nodes into functions.}
    \label{fig:exprtree}
    \Description{This figure shows a diagram and two programs. The diagram is a tree representing an arithmetic expression. The two programs are versions of the tree turned into code. The middle one contains variables like "k = 4" and the right one contains functions like "def x()".}
\end{figure*}


When tracing a program, a person needs to know what instruction to execute next, i.e. how to follow the program's control flow. If a person adopts a linear tracing strategy in a straight-line program, they only need to know the ``instruction pointer'', or the current position in the program. A person can trivially determine whether a line has been previously executed (above the pointer), or still needs to be executed (below the pointer). 

In what situations, then, does a person need to remember more about control flow than the instruction pointer? Consider the three equivalent representations of an arithmetic expression shown in Figure~\ref{fig:exprtree}. A trace through these programs is akin to a walk around the expression tree. For example, linearly tracing the variable program in Figure~\ref{fig:exprtree} starts from the leaves: observing $1$ and $4$, then computing $1 - 4$, and so on. By contrast, an on-demand tracing strategy for the variable program starts at the root: observing that the goal is \verb|u - s|, then computing \verb|u = v + j|, and so on. A standard tracing strategy for the function program should look similar to the on-demand tracing strategy for the variable program: start with \verb|x()|, observe that \verb|x| depends on \verb|u() - s()|, then compute \verb|u()| and so on.

Because the variable program must be a topological sort of the tree, tracing linearly ensures that an expression's dependencies have always been computed before computing the expression. However, this property is lost if the person adopts a non-linear tracing strategy, or if the program itself is non-linear. When tracing the variable program on-demand or tracing the function program, a person must maintain a set of previously visited program positions (i.e. a \textit{call stack}). After visiting $\verb|x()| \rightarrow \verb|u()| \rightarrow \verb|v()|$, the tracer must remember that \verb|v()| was contained in the definition of \verb|u|. This information must either be retrieved from the program text or stored in working memory.

In the abstract, the on-demand tracing strategy has a task structure where a goal (e.g. compute \verb|x = u - s|) and its state (\verb|u = 13|) become temporarily replaced by a sub-goal (compute \verb|s|) with separate state. Within cognitive load theory, this hierarchical subgoal-within-goal task structure has been consistently shown to induce working memory errors:

\begin{itemize}
    \item Students solving basic Lisp programming problems make more errors attributable to working memory when the problem uses a more deeply nested expression\,\cite{anderson_novice_1985}.
    \item Students solving multi-step geometry problems (e.g. ``to compute angle X, that's 180 - Y, so now I compute Y'') make errors most commonly within sub-goal stages where WM load is highest\,\cite{ayres_locus_1990}. 
    \item Students mentally distributing multiplication over algebraic expressions (e.g.\ $-3(-4 - 5x) - 2(-3x - 4)$) make errors most commonly when expanding the the second parenthetical, because the state of the first expansion must be held in WM while accomplishing the secondary goal\,\cite{ayres2001systematic}. 
\end{itemize}

Hence, we should expect to people make increased WM errors when tracing a program with  nested control flow.

\section{Experiments}

We ran four experiments to explore the role of WM for tracing with two kinds of program state: data (variable/value pairs) and control flow (tracking data dependencies).

\subsection{Variable recall}

First, we sought to measure the total capacity of WM for variable/value pairs without any interference. This provides a baseline for later comparing how other tracing tasks (e.g. arithmetic) affect WM capacity. Specifically, we asked two questions:

\begin{enumerate}
    \item How many variable/value pairs can a person keep in WM?
    \item How does the kind of variable name affect its memorability?
\end{enumerate}

In a realistic program, both of these questions are inevitably confounded by the semantic relationship between a variable and its value. For example, \verb|greeting = "Hello"| is likely easier to remember than \verb|foo = "Hello"| because ``Hello'' is a kind of greeting. To avoid this confounder, we will consider the case where a variable has no particular relationship to its value. We only consider numeric values, and we consider variable names that are either common English nouns or single letters.

For the first question, our hypothesis is that participants should hold about 5 variable/value pairs in WM, consistent with prior work on cued recall in paired-associate learning\,\cite{nobel2001retrieval,unsworth2009examining}. For the second question, we hypothesize that English words should be more memorable than single letters due to having \textit{some} meaning if not a \textit{relevant} meaning, and therefore should correspond to more pairs remembered on average.

\subsubsection{Methodology}

In this experiment, we tested WM capacity via cued recall. Participants were presented with several expressions of the form \verb|variable = value| for 2s each. They were then prompted to answer \verb|variable = ?| for every observed pair. The specific methodology and parameters are similar to those used in Nobel and Shiffrin's paired-associate experiments\,\cite{nobel2001retrieval}. Within a given trial, we randomly generated 10 variable/value pairs. Each value is a digit from 0 to 9. We have two different conditions for variable names: single-letters (A to Z), and common English nouns (e.g. cave, tax, cherries). The participant was presented with one pair at a time for 2s per pair. The presentation used programming syntax, e.g. \verb|x = 4|. After the final pair, the participant was prompted to input the corresponding value of all variables, e.g. \verb|x = ?; f = ?|. The order of prompts was randomized with respect to the order of initial presentation. (The randomization was used to defeat mnemonic strategies that simply memorized variables and values as two separate streams of serial data which we observed in a pilot.) We then measured the participant's accuracy as the number of values correctly recalled.

Unlike Nobel and Shiffrin, we did not add a distractor task (e.g.\ mental arithmetic) inbetween the presentation and prompt phases. Their goal was to measure long-term memory capacity, while our goal is to measure WM capacity. As a result, we may expect to see a larger number of recalled pairs than in their experiment.

Each participant completed 4 trials per condition. The order of trials was counterbalanced, and this experiment had 15 participants. In this and all other experiments, we recruited participants with the following criteria and instructions:

\begin{itemize}
    \item \textbf{Participant pool:} Participants were recruited from Amazon Mechanical Turk within the United States and Canada. They were compensated a fixed amount, estimated by timing the average duration of a pilot study and paid a corresponding amount prorated at \$15/hr. 
    \item \textbf{Required expertise:} We required that participants have a basic competency with tracing Python programs, which we ensured with a short pre-test before each experiment. Participants also needed to correctly complete a sample trial before doing an experiment.
    \item \textbf{No memory aids:} Participants were instructed to not use pen/paper or any other external media as a memory aid.
    \item \textbf{No distractions:} Participants were asked to complete the experiment without any visual or audio distractions, and to complete the experiment in one sitting.
\end{itemize}

\subsubsection{Results}

\begin{figure}[t]
    \centering
    \includegraphics[width=\columnwidth]{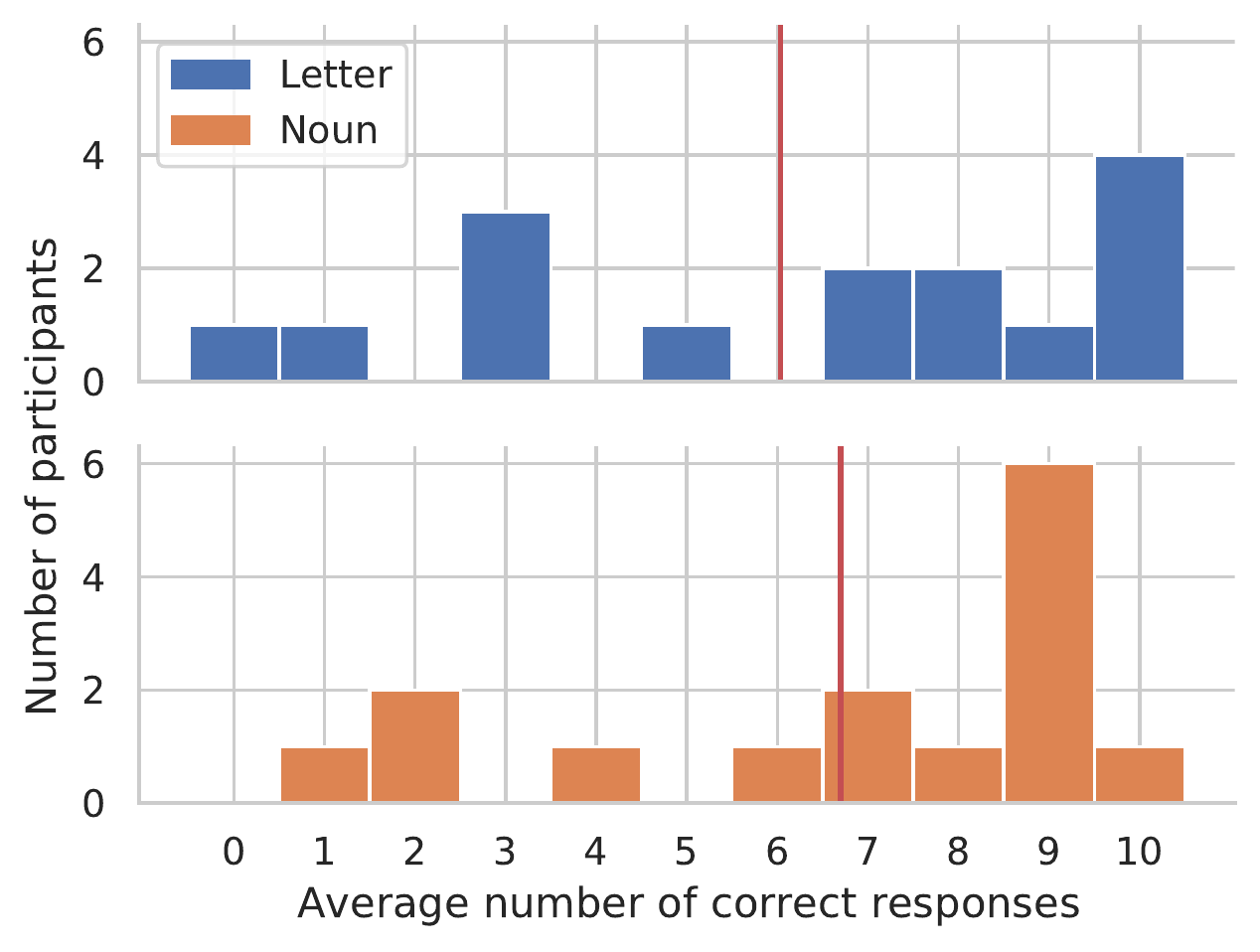}
    \caption{Number of values correctly recalled when cued with the paired variable, averaged by participant. The red line indicates the mean value of each distribution.}
    \label{fig:exp1}
    \Description{The left column shows histograms with the average number of correct responses on the x-axis and number of participants on the y-axis. The top histogram represents the letter condition, and the bottom histogram represents the noun condition. In both histograms, many participants got 9 or 10 responses right with a long tail towards the left. The right column shows the same histograms but cumulative. }
\end{figure}

The distribution of average accuracy by participant in each condition is shown in Figure~\ref{fig:exp1}. On average across both conditions, participants were able to recall a mean/standard deviation of $6.5$ ($\sigma = 3.7$) for letters and $7.2$ ($\sigma = 3.4$) for nouns, and a median of 7.5/9.0. Notably, 5 out of 15 participants achieved an average of 9 or better in both conditions. Using a Wilcoxon signed-ranks test, the difference in accuracy between the conditions was not statistically significant ($T = 40.5, p = 0.45$).

\subsubsection{Discussion} For semantically-unrelated variable/value pairs, these results suggest that a person can remember on average about 7 pairs in a pure memorization setting. This number is higher than hypothesized, likely because our experiment did not include a distractor tasks like the related work. The notion of average capacity may also be somewhat tenuous if memory capacity varies so widely across individuals. Participants could not remember significantly more pairs when random nouns were used as variable names. This result suggests that nouns are not more memorable, or name memorability does not matter if the semantic relationship of variable and value is arbitrary.

\subsection{Variable recall with arithmetic}

Program tracing involves the maintenance of program state while performing operations like arithmetic. So next, we consider: how does WM capacity for variable/value pairs change when interleaved with mental arithmetic? Within the multi-component model of WM\,\cite{baddeley_working_1992}, prior work has found that three-digit mental arithmetic interferes with simultaneous tasks that use central executive WM, but not phonological WM\,\cite{furst2000separate}. Single-digit mental arithmetic problems have also been repeatedly confirmed to use central executive WM\,\cite{destefano_role_2004}. The cognitive action of updating WM in a running memory task is coordinated by central executive WM\,\cite{morris1990memory}, therefore we hypothesize that participants should have a lower effective WM capacity for variable/value pairs than in the prior experiment.

\subsubsection{Methodology} 

In this experiment, participants traced a straight-line program of variables assigned to arithmetic expressions, one line at a time and without the ability to look back. We measured WM capacity by the point at which the participant provides an incorrect response. Specifically, we randomly generated programs of the form:

\begin{lstlisting}[language=Python]
x = 3 + 4
t = x - 1
b = t + x
z = x - b
# ...and so on 
\end{lstlisting}

Each line assigns a new variable to an arithmetic expression that involves randomly selected variables from the previous lines (or a constant for lines 1-2). Variables names are lower-case single letters and all constants are between 0 and 9. The only arithmetic operations are addition and subtraction to keep the mental effort of any individual operation relatively small. The program is 11 lines long, such that the participant would have to remember 10 variables by the last line.

Within a given trial, we present the participant one line at a time. In the above example, the participant would start by seeing: ``\verb|x = 3 + 4|. What is the value of \verb|x|?'' After entering 7, that line disappears and the participant sees: ``\verb|t = x - 1|. What is the value of \verb|t|?'' This way, the participant must commit the variable/value pairs to working memory, instead of looking them up later. After entering a value, the software waits for two seconds before proceeding to the next line. This process repeats until the participant responds incorrectly.

An incorrect response cannot necessarily be attributed to a working memory failure. The participant could simply add two numbers incorrectly (which could itself be a working memory failure, but for now we ignore that possibility). Using the methodology of Campbell and Charness two-digit squaring experiment\,\cite{campbell_age-related_1991}, we classify the incorrect response as either a substitution error or calculation error. Substitution means the participant entered an a number that could have been computed by mixing up two previously seen variables. For example, if the current state is \verb|x = 7; t = 6; b = 13| and for \verb|x - b| the participant enters \verb|1|, then we classify that as a substitution of \verb|t| for \verb|b|, a failure of working memory. All other incorrect numbers are considered calculation errors.

Each participant completed 10 trials, and this experiment had 15 participants (one participant was excluded for on average failing at the first line, leaving 14 participants).

\subsubsection{Results}

\begin{figure}[t]
    \centering
    \includegraphics[width=\columnwidth]{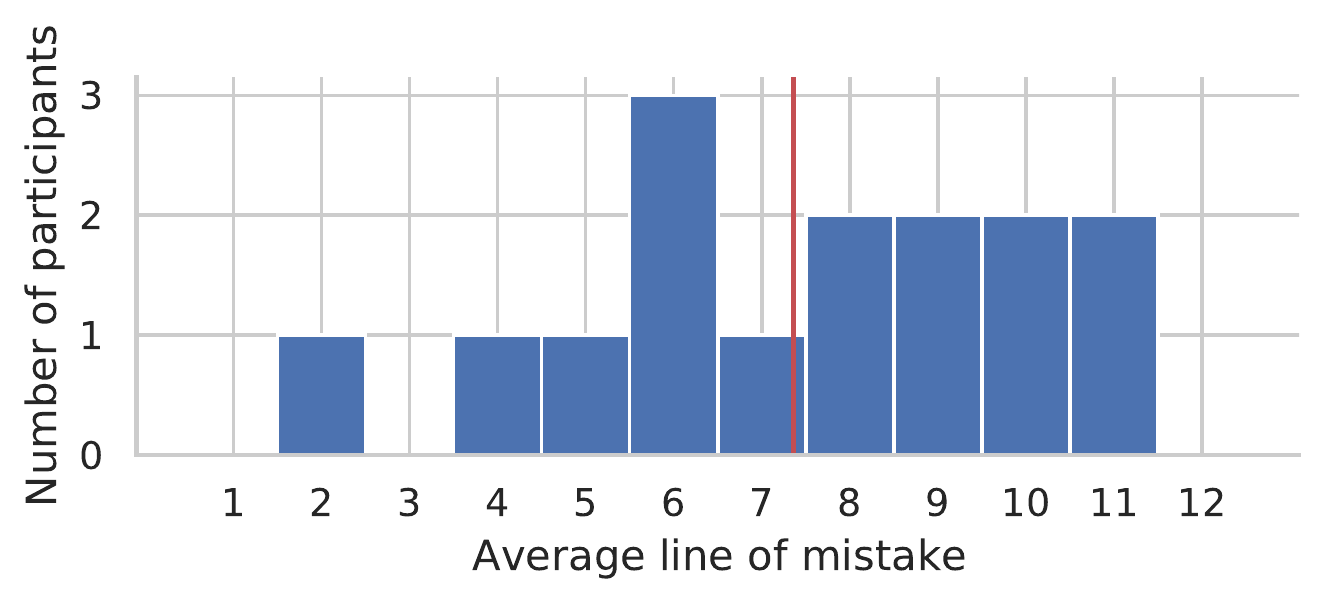}
    \caption{Histogram of the average line of code at which a participant made a mistake in Experiment 2. }
    \label{fig:exp2_results}
    \Description{The graph is a histogram of the average final line per participant. Participant averages range from 4 to 11, with a median of 8.}
\end{figure}

Each participant finished the experiment at a particular line, either by providing an incorrect response (final line 1-11) or completing the task without error (final line 12). The histogram of participants' average final line is shown in Figure~\ref{fig:exp2_results}. Across all participants, the average final line was 7.9 ($\sigma$ = 3.7) and median was 8.0.

To compare these results to Experiment 1, first we have to model WM capacity in terms of the line of failure. If participants failed at line 8, then they either calculated incorrectly or forgot at least one of 7 preceding variables, suggesting a WM capacity of 6 variables. Hence, we model effective WM capacity as $\text{line number} - 2$, indicating the average WM capacity for this experiment was 5.9. To test the hypothesis of whether mental calculations reduced WM capacity, we compared the distribution of WM capacity in Experiment 2 vs. the distribution of WM capacity measured in Experiment 1 in the letter condition. Using a Kruskal-Wallis test, the difference was not significant $(H(1) = 1.51, p = 0.21)$.

\begin{figure}[t]
    \centering
    \includegraphics[width=\columnwidth]{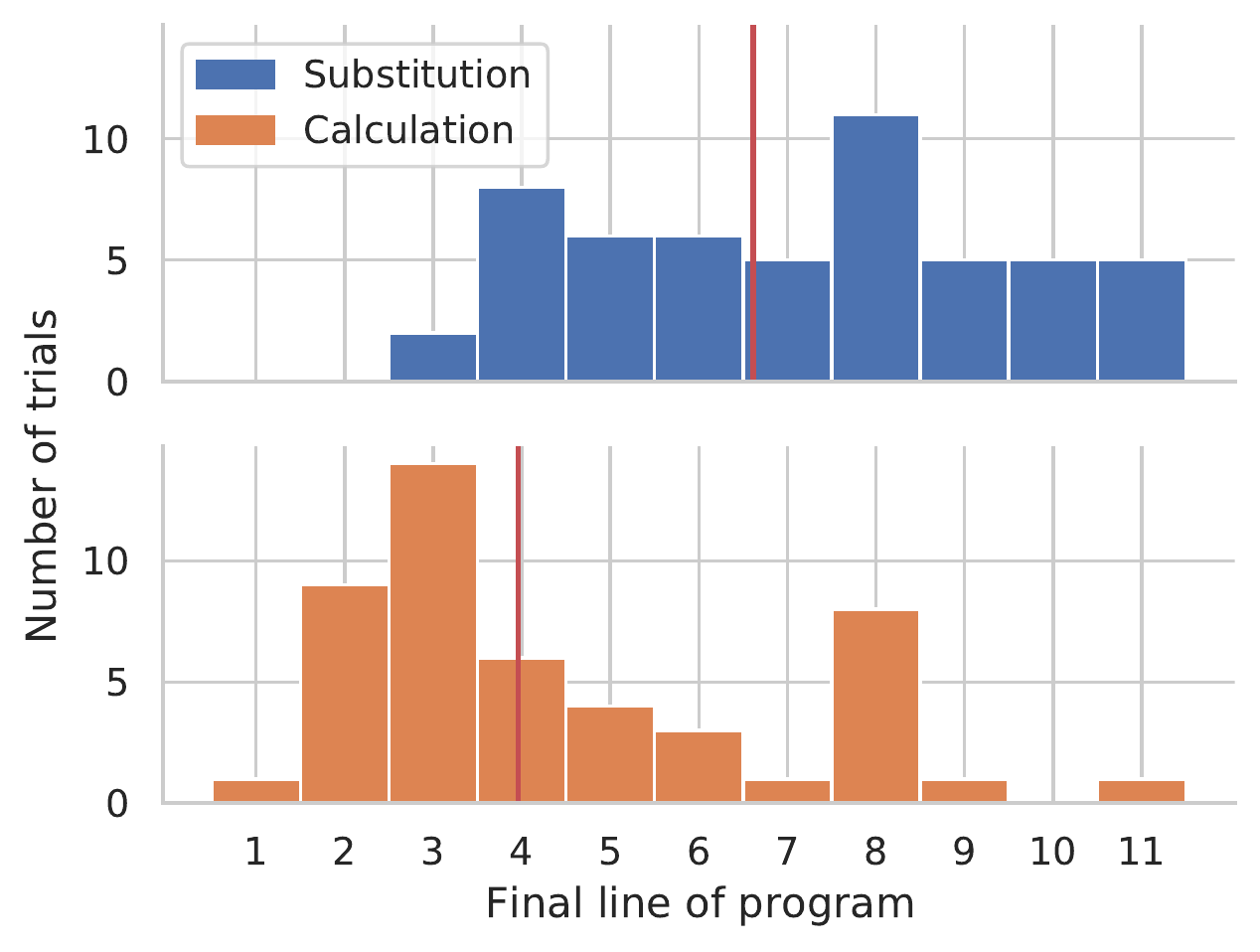}
    \caption{Histogram of error types at each line in Experiment 2. Participants made 70\% more substitution errors than calculation errors.}
    \label{fig:exp2_errors}
    \Description{The graph is a histogram of the kind of errors made at each line. Most calculation errors happened in lines 2 to 8 while most substitution errors happened in lines 4 to 11.}
\end{figure}

For each trial where the participant gave an incorrect response, we classified the error as substitution or calculation. The distribution of errors by type is shown in Figure~\ref{fig:exp2_errors}. Of the 101 total errors, 53 were substitution and 48 were calculation. The median calculation error happened at line 3.5, while the median substitution error happened at line 7.0. Using a Kruskal-Wallis test, the difference was significant ($H(1) = 25.57, p < 0.001$).

\subsubsection{Discussion}

These results do not support the hypothesis that performing mental arithmetic reduces WM capacity. Participants in Experiment 2 on average remembered fewer variables than Experiment 1, but the different is not statistically significant. However, the experimental designs are somewhat different (e.g. continually prompting for information vs. asking for it all at once) and the test is between-subjects, so future work can control for these factors. Consistent with Campbell and Charness\,\cite{campbell_age-related_1991}, we found that over half of mistakes can be attributed to substitution in WM. Specifically, errors could be explained by accidentally swapping the association between variables in WM, as opposed to forgetting them entirely.

\subsection{Straight-line tracing strategy}

Next, we consider how WM influences tracing for straight-line programs when the complete program is available at all times. We investigate two questions:

\begin{enumerate}
    \item Do participants demonstrate tracing strategies that diverge from linear execution?
    \item Where in the program are participants most likely to have working memory errors? 
\end{enumerate}

As described in Sections 2.3 and 2.4, we expect there to be two basic tracing strategies: linear and on-demand. Prior work has shown that people will trace somewhat out of linear order\,\cite{vainio_factors_2007}, but we do not know precisely to what extent or how often. Hence our first goal for this experiment was to design a code viewing interface such that we can deduce the participants' strategies from their viewing patterns. Then once we have identified which strategy a participant is using, we can hypothesize where errors are likely to occur:
\begin{itemize}
    \item For linear tracing, the primary source of WM load should be remembering variable/value bindings. Based on Hitch's model of errors in mental arithmetic\,\cite{hitch_role_1978}, a person is more likely to forget values computed earlier in the program than later. Then participants should forget computations in the top lines of the program (leaf nodes) more frequently than later lines (inner nodes).
    \item For on-demand tracing, the primary source of WM load should be tracking the current path in the expression tree. Based on the predictions of cognitive load theory for hierarchical task structures\,\cite{anderson_novice_1985,ayres_locus_1990,ayres2001systematic}, participants should make the most WM errors at the deepest sub-goal (i.e.\ leaf nodes). A WM error would cause a participant to forget their path to the leaf (i.e.\ inner nodes), so participants should revisit inner nodes more frequently than leaf nodes.
\end{itemize}

\subsubsection{Methodology}

\begin{figure}[t]
    \centering
    \includegraphics[width=0.8\columnwidth]{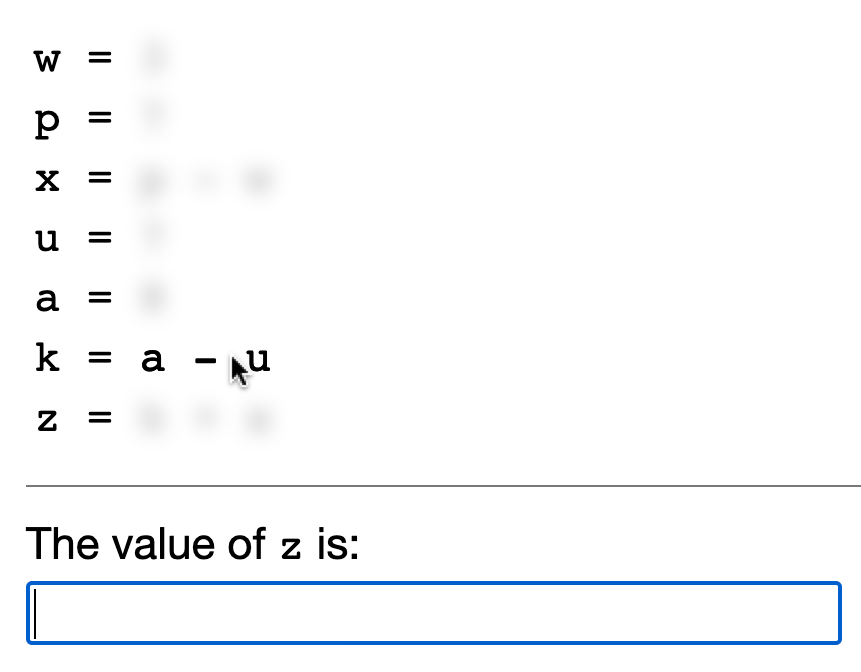}
    \caption{Interface for tracking the participant's attention during tracing. Each expression is blurred unless the particpant's cursor hovers over it.}
    \label{fig:blurry_ide}
    \Description{a screenshot of an interface. Shows a program that's a sequence of variables assigned to arithmetic expressions with the expressions blurred out, except for one that the screenshotted mouse is hovering over. Beneath the sequence is an input box that is labeled "The value of z is"}
\end{figure}

In this experiment, we track where a participant's attention goes as they trace through a straight-line program. Given the participant's attention over time, we classify their tracing strategy as linear or on-demand. Then we identify working memory errors in their traces based on re-visits to lines of the program.



To track attention during tracing, we created a code viewing interface where each expression is blurred by default, i.e. a restricted focus viewer\,\cite{jansen2003tool} (we did not use an eye-tracker so the experiment could be performed in a browser on Mechanical Turk). The participant can move their mouse over an individual line to bring it into focus. This way, the mouse acts like a foveal region in an eye tracking experiment --- it represents the user's current focus. W Figure~\ref{fig:blurry_ide} shows the interface. Because the mouse may briefly hover over other lines while moving to a destination, we discard any hover actions that occurred for less than 300ms. This number was selected by inspecting the noise in the data, but in the spirit of multiverse analysis\,\cite{steegen2016increasing} we confirmed that all significant results still hold with $p < 0.05$ for a threshold of 100ms and 500ms.

\begin{figure}[t]
    \centering
    \includegraphics[width=\columnwidth]{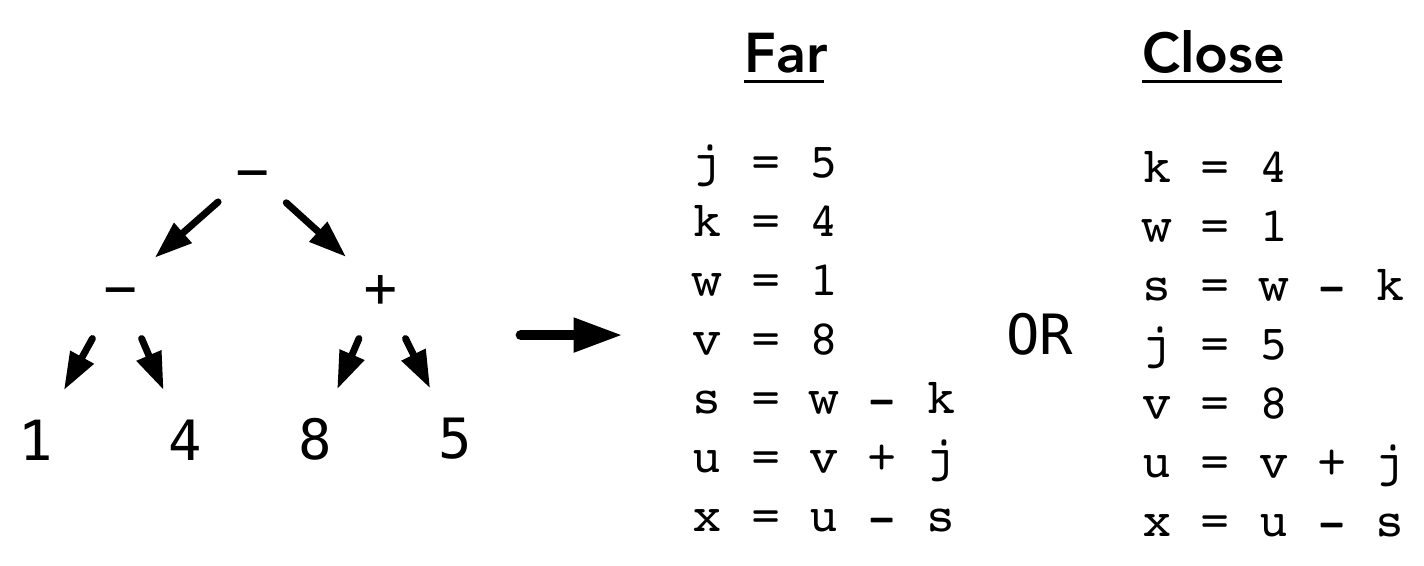}
    \caption{Process of turning an expression tree into a program. Far has the highest average distance between variable definition and use, while Close has the lowest.}
    \label{fig:tree_to_straightline}
    \Description{this diagram shows an arithmetic expression tree turning into two programs. Both are topologically sorted with respect to the tree, but one has a higher average distance between variables than the other.}
\end{figure}

To generate straight-line programs, we start by randomly generating an arithmetic tree of size 7 with constants at the leaves and binary operators at the inner nodes. Every node is given a random single-letter variable name. To map a tree to a straight-line program, we have to pick an ordering that respects the dependencies in the tree, i.e. any topological sort. The main difference between sorts is the distance from a variable definition to its usage. For this experiment, we consider two conditions: sorts with the highest average distance (``Far''), and lowest average distance (``Close''). Figure~\ref{fig:tree_to_straightline} shows an example translation.


Each participant completed 5 trials in each condition (Far/Close), and this experiment had 15 participants. 3 participants were removed due to exceptionally poor performance (accuracy less than 20\%), leaving 12 participants.

\begin{figure*}[t]
    \centering
    \includegraphics[width=\textwidth]{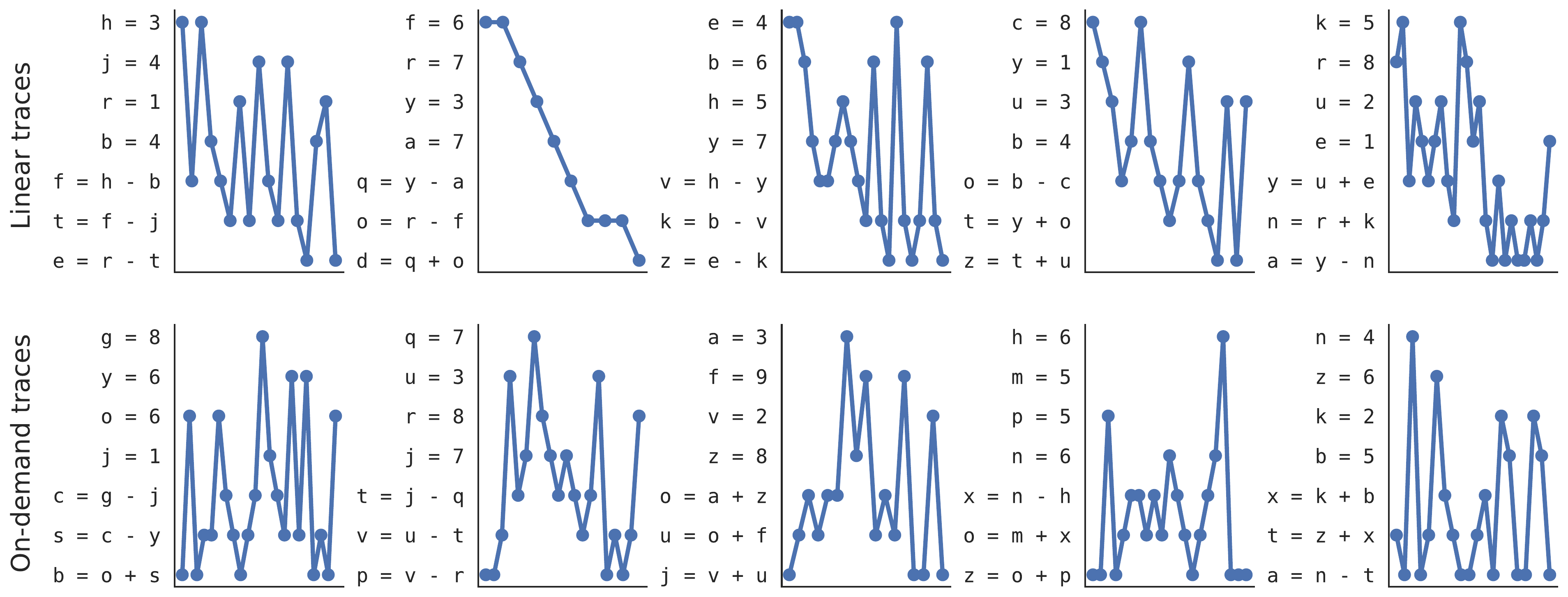}
    \caption{Example traces of straight-line programs categorized by strategy. The x-axis is time, and the y-axis is the lines of the program. Each dot is a hover event.}
    \label{fig:exp3_examples}
    \Description{ten examples are shown of participants' traces through straight line code. The linear traces start at the top line and go towards the bottom, while the on-demand traces start at the bottom line.}
\end{figure*}

\subsubsection{Results}

\begin{figure}[t]
    \centering
    \includegraphics[width=\columnwidth]{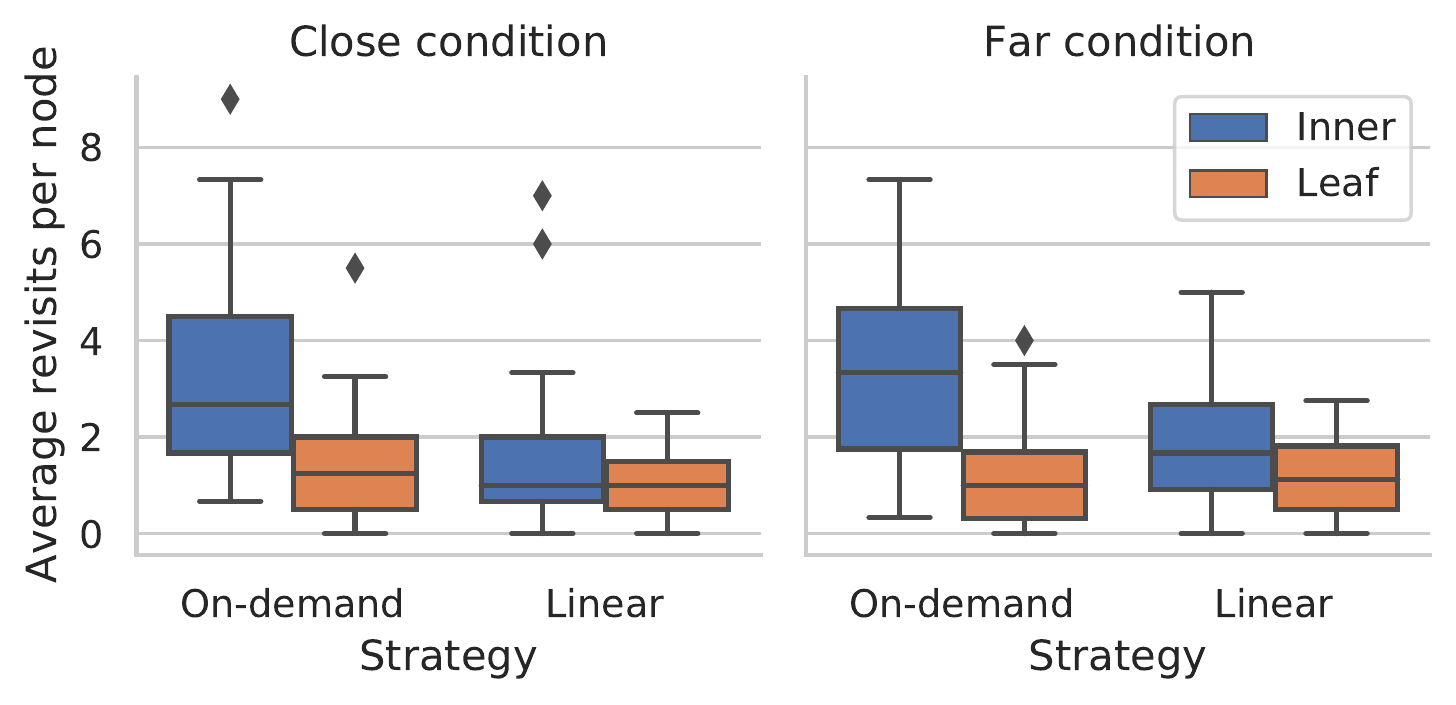}
    \caption{Distribution of average revisits to expression tree nodes, separated by node type, tracing strategy, and program type. The difference between visits to inner nodes and leaves was greater in the on-demand strategy than the linear strategy.}
    \label{fig:exp3_results} 
    \Description{Two graphs containing boxplots of average revisits split by three factors: variable distance, strategy, and node type. The two outer graphs have about the same distribution. Within each graph, the on-demand strategy has a higher average for inner nodes than leaf nodes, while the linear strategy has them about the same.}
\end{figure}

First, for each trial, we classified the participant's mouse movements as indicating an on-demand or linear strategy. As a simple heuristic, we classify a trial as linear if the first line is first visited before the last line, and on-demand otherwise. Across all trials, 55\% were classified as on-demand and 45\% as linear. Most participants appeared to individually prefer one strategy over another --- 9 out of 12 participants adopted a single strategy at least 70\% of the time. Figure~\ref{fig:exp3_examples} shows a sample of actual traces classified as linear (top) and on-demand (bottom) in the Far condition. In these examples, we can already observe a few WM influences:
\begin{itemize}
    \item The top-left and top-center participants with the linear strategy traced linearly in the set of binary operations, but looked up constant values on-demand (consistent with Vainio and Sajaniemi\,\cite{vainio_factors_2007}). Others like the top-middle-left participant traced linearly in both variables and binary operations.
    \item Some actions are clearly attributable to strategy vs. WM error. For example, the bottom-left participant goes from \verb|b = o + s| to \verb|o = 6|, indicating an intentional on-demand shift in attention, but then returns to \verb|b| before moving to \verb|s = c - y|. This indicates a WM error of forgetting the second operand to \verb|b|. By contrast, the bottom-middle-left participant goes straight from \verb|u| to \verb|t| after observing \verb|v = u - t|, indicating no comparable WM error.
\end{itemize}

To quantitatively analyze forgetting, we counted the number of times a participant re-visited each line of the program. We then categorized the lines as either leaves of the expression tree (constant variables) or inner nodes (binary operations). Figure~\ref{fig:exp3_results} shows the distribution of average re-visits separated by participant's strategy and the program sorting condition. To test for differences between distributions, we fit a linear mixed model with a three-way interaction of strategy, node type, and variable distance as the fixed effects and participant as the random effect. The number of revisits was the dependent variable. A one-way ANOVA showed a significant difference between revisits in the interaction of node type and strategy ($p < 0.001$), but not variable distance. We ran six post-hoc T-tests on all contrasts between pairs of the interaction, adjusted by the Tukey method. We found:

\begin{itemize}
    \item Participants on average revisited nodes 1.37 more times per node in the on-demand strategy than the linear strategy ($T(200) = 8.66, p < 0.001$).
    \item Participants revisited inner nodes more than leaf nodes on average 2.1 more times in the on-demand strategy ($T(200) = 9.95, p < 0.001$), and 0.63 times in the linear strategy ($T(200) = 2.69, p = 0.038$).
    \item Participants revisited inner nodes an average of 1.11 more times in the the on-demand strategy than the linear strategy ($T(216) = 4.23, p < 0.001$).
\end{itemize}

\subsubsection{Discussion}

We have found clear evidence that participants will adopt both on-demand and linear strategies for tracing straight-line code. The results also support our hypothesis that WM theory predicts where a participant will make errors given a strategy: at leaves for linear, and at inner nodes for on-demand. Moreover, we also found evidence that the linear strategy causes fewer WM errors than on-demand. That suggests that the cognitive load caused by tracking paths through the tree is greater than storing variable/value pairs in WM.

\subsection{Function tracing strategy}

Finally, we extend the methodology of the prior experiment to programs with functions. Similarly, we want to understand: what strategies will people use to trace programs with functions, and where are they most likely to make errors? Recall from Section 2.4 that tracing a function program starting from the top-level function call is equivalent to an on-demand tracing strategy in a straight-line program. When projected onto the abstract expression tree, both traces start at the root and proceed to the leaves. 

However, now a linear strategy should be harder than before, because a person has to mentally reconstruct the dependency graph and topological sort before tracing linearly. Hence we hypothesize that more people should adopt an on-demand strategy than in the prior experiment. For a given strategy, though, we hypothesize that WM errors should occur most frequently as predicted in the prior experiment: on-demand traces forget the inner nodes, and linear traces forget the leaves.

\subsubsection{Methodology} 

\begin{figure}[t]
    \centering
    \includegraphics[width=\columnwidth]{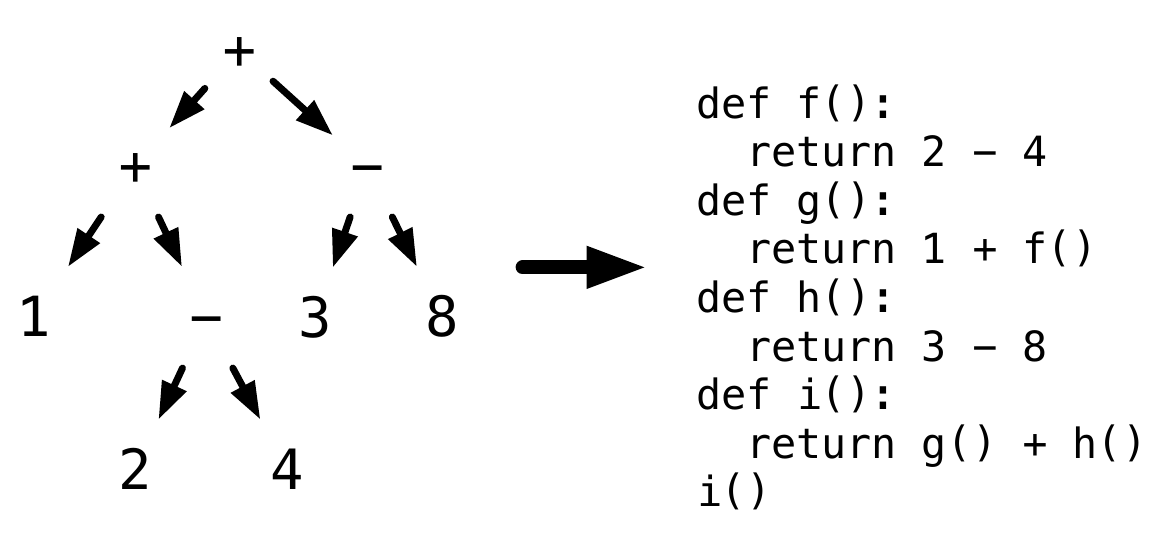}
    \hspace{0.03\textwidth}
    \caption{Transformation of an expression tree into a sequence of functions. We only convert binary operations into functions, because turning every node into a function took too much per-trial time for participants in a pilot study.}
    \label{fig:tree_to_func}
    \Description{The diagram shows an arithmetic expression tree turning into a program with four straight-line definitions of functions, e.g. \verb| def f(): ...|.}
\end{figure}

\begin{figure}[t]
    \centering
    \includegraphics[width=0.8\columnwidth]{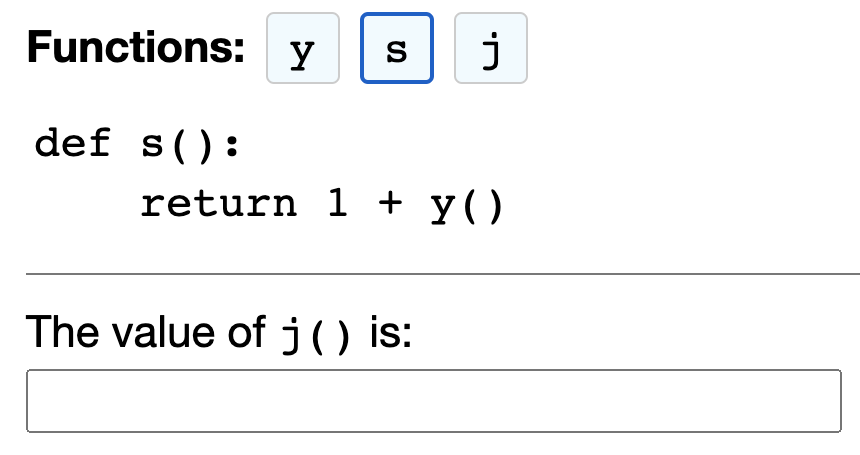}
    \caption{Experiment interface for showing participants one function at a time.}
    \label{fig:exp4_interface}
    \Description{The screenshot shows the experiment interface. At the top, it contains a bank of buttons for each function. In the middle is the code for a single function. At the bottom is an input box prompting for the value of one of the functions.}
\end{figure}

In this experiment, participants trace through a program with functions while we track which function has their attention. Like the prior experiment, we randomly generate an expression tree and convert it into a program with functions as shown in Figure~\ref{fig:tree_to_func}. Then to track the participant's attention, we created a code viewing environment where one function is displayed at a time, shown in Figure~\ref{fig:exp4_interface}. The participant is given a bank of buttons with the name of each function, and clicking on the button goes to the function's source, similar to an IDE. A participant's trace through the expression tree corresponds to their sequence of button clicks. The order of buttons is randomized, so it bears no relationship with respect to the functions' order of usage in the program.

Each participant completed 10 trials, and this experiment has 15 participants (1 was removed for poor performance, leaving 14 participants).

\subsubsection{Results}

\begin{figure*}[t]
    \centering
    \includegraphics[width=0.95\textwidth]{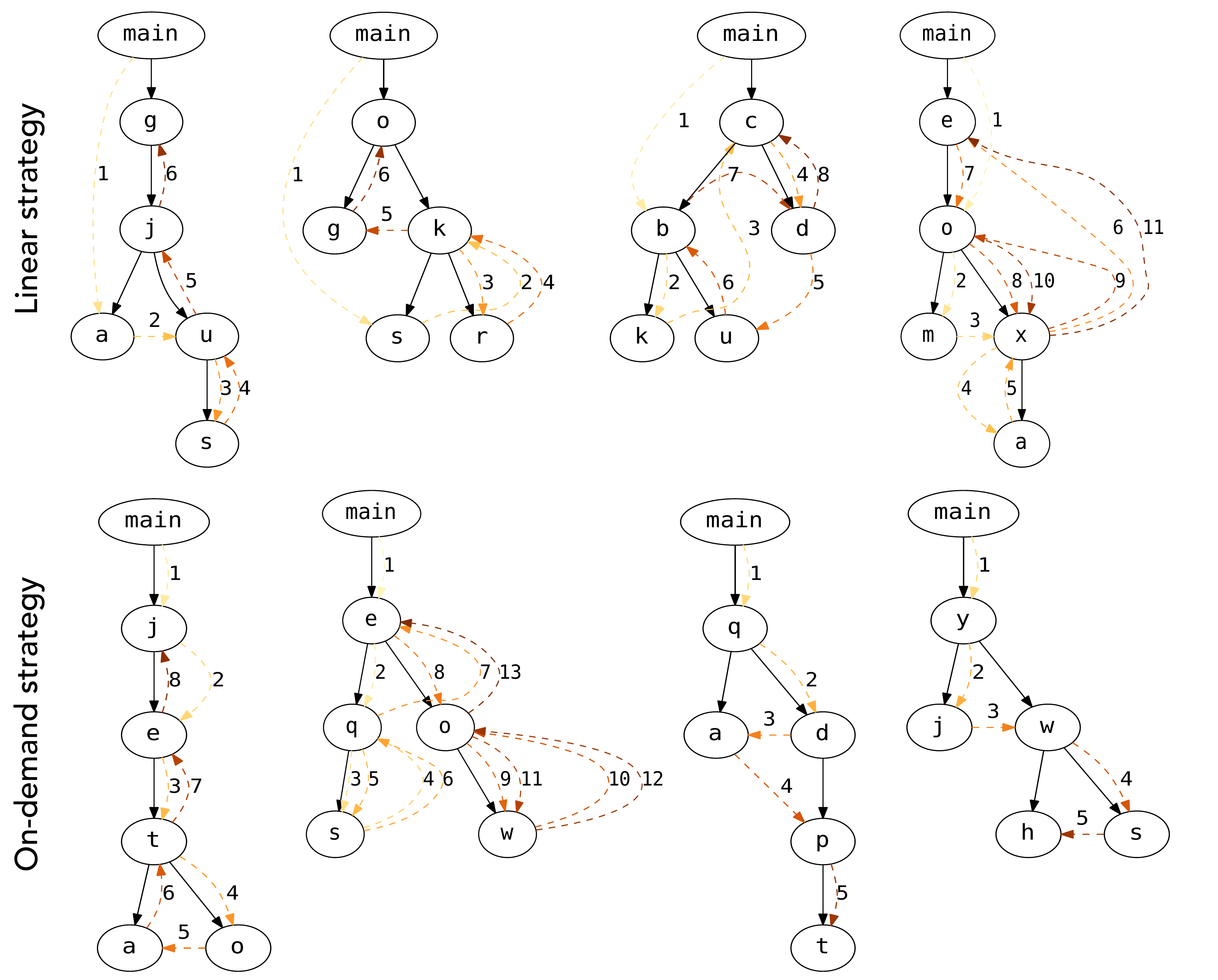}
    \caption{A sample of traces from the participants, grouped by strategy. Each tree represents a dependency graph of function calls e.g. \protect\Verb|j| $\rightarrow$ \protect\Verb|e| means the definition of \protect\Verb|j| calls \protect\Verb|e()|.}
    \label{fig:exp4_examples}
    \Description{This figure shows eight examples of traces through the function dependency tree. Four are for the linear strategy, where the participant did not go to the root function node first. Four are for the on-demand strategy, where the participant did go to the root function node first.}
\end{figure*}

First, we classified each trace as an on-demand or linear strategy. We used a simple heuristic similar to before: if the participant visited the root function node first, they used an on-demand strategy, and they used a linear strategy for visiting any other node first. Figure~\ref{fig:exp4_examples} shows several examples of traces around the expression tree within each strategy. Out of 140 trials, 54\% used a linear strategy and 46\% used an on-demand strategy. Like the prior experiment, 11/14 participants consistently picked one strategy at least 70\% of the time.

\begin{figure}[t]
    \centering
    \includegraphics[width=0.8\columnwidth]{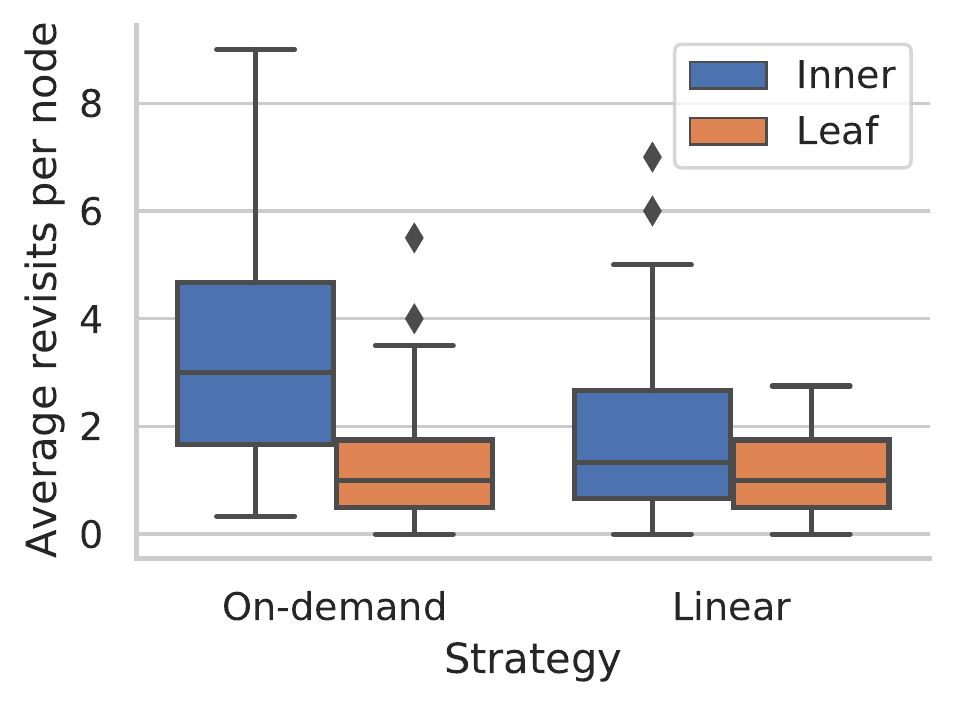}
    \caption{Distribution of revisits to nodes by node type and strategy. Participants visited more inner nodes than leaf nodes in the on-demand strategy, but not in the linear strategy.}
    \label{fig:exp4_results}
    \Description{A boxplot graph of the average revisits in Experiment 4. It shows four boxplots, grouped by strategy and node type. The on-demand strategy has a higher average for revisiting inner nodes than leaf nodes, while the linear strategy has a lower average for inner nodes than leaf nodes.}
\end{figure}

Next, we computed the average number of revisits to inner and leaf nodes in each trace. The distribution of revisits by strategy and node type is shown in Figure~\ref{fig:exp4_results}. As in the previous experiment, we fit a linear mixed model with strategy and node type as fixed effects and the participant as a random effect, and the number of revisits as the dependent variable. Using a one-way ANOVA, the only statistically significant difference in revisits is the node type $(F(263) = 41.06, p < 0.001)$.
\subsubsection{Discussion} 

These results provide mixed support for our first hypothesis. It was not true that most participants preferred an on-demand strategy. In fact, we found the preponderance of linear traces somewhat bewildering --- participants appeared to just click through each function, remembering what they could and then synthesizing the information into an answer at the end. However, the results do suggest that the linear strategy had increased WM load for function programs compared to straight-line programs. In Experiment 4, there was not a statistically significant difference between WM errors in the on-demand and linear strategies, whereas on-demand traces contained more errors in Experiment 3.

The results do not support our second hypothesis, that participants revisit nodes as predicted by WM theory. The relative revisits to inner vs. leaf nodes was not different between strategies. One possible explanation is that the notion of a linear strategy is different for the straight-line interface than the function interface. The function interface does not expose the call graph, so a participant cannot easily walk from the leaves to the root as they can when scanning top-to-bottom in the straight-line code.


\section{General Discussion}

\begin{figure*}[t]
    \centering
    \includegraphics[width=\textwidth]{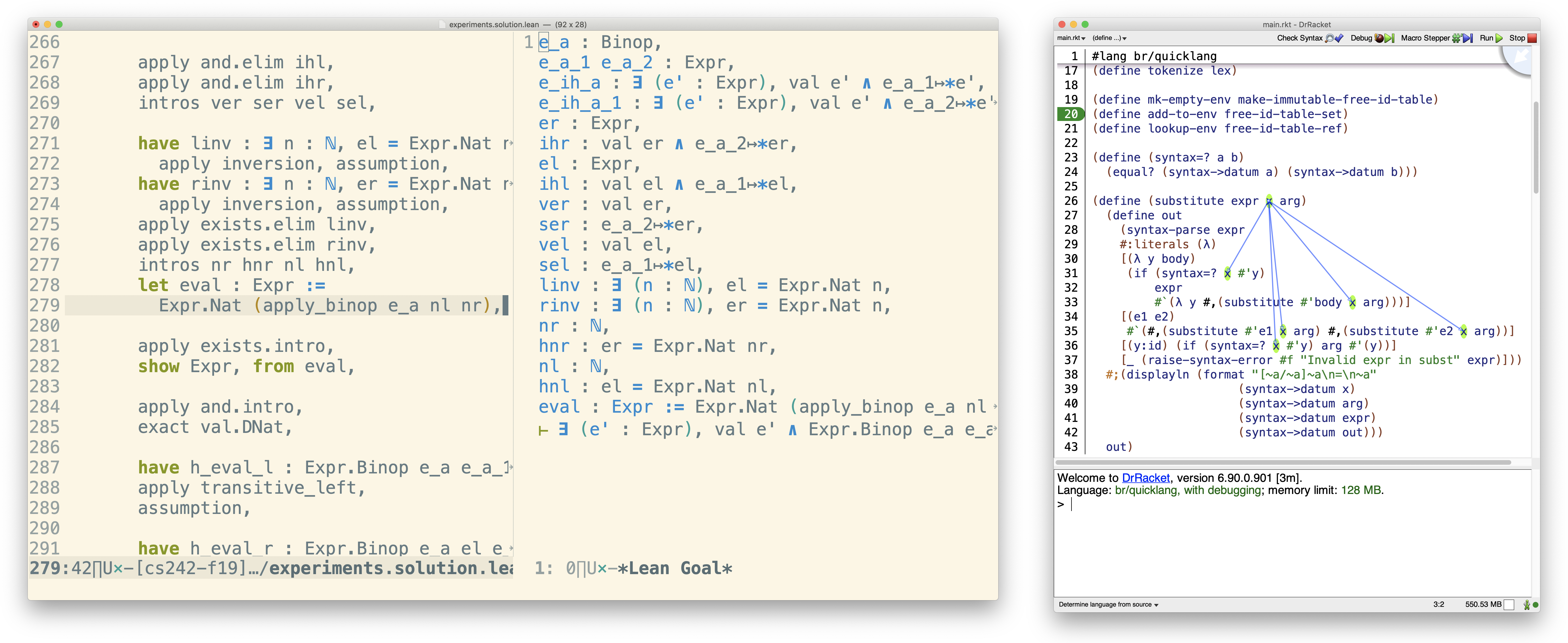}
    \caption{Left: the Lean theorem proving language inside its Emacs IDE. The right pane shows all variables in scope on the current line in the left pane. Right: the DrRacket IDE shows all usages of a variable on mouse hover.}
    \label{fig:ides}
    \Description{The left screenshot shows an IDE inside Emacs. It is split into two windows. On the left window is program code, and on the right window is a list of variables. The right screenshot shows the DrRacket IDE. It contains a program where one variable is highlighted, with lines drawn to the other program locations referencing that variable.}
\end{figure*}

In summary, we made six findings from our experiments:

\begin{itemize}
    \item A person can hold about 7 variable/value pairs in working memory, and can remember variables named as random nouns equally well as random letters.
    \item Single-digit mental arithmetic does not reduce WM capacity for variable/value pairs.
    \item When recalling values for mental arithmetic, a majority of errors could be explained by recalling a value bound to a different variable held in WM.
    \item People will trace a straight-line program both linearly and on-demand, and a given person is likely to stick with one strategy.
    \item People will make WM errors in different parts of the program based on their tracing strategy, and on-demand strategy causes more WM errors than the linear strategy for a straight-line program.
    \item People will not necessarily trace a function in order of execution.
\end{itemize}

These findings occurred within a controlled experimental setting and within a specific application domain. As with all laboratory / cognitivist research, this naturally raises the question --- do these results generalize to realistic programs? And how can we practicably apply these results to high-level design?

\subsection{Limitations}

Our findings are not universal statements about e.g. all variable/value pairs. For example, if a person tries to remember \verb|x = [1, 5, 7]|, the list is a more complex object to be stored in WM than a single digit. A person likely cannot remember as many variable/list pairs as variable/number pairs. More generally, several factors limit the generality of our current results and point to interesting future work. %
\begin{itemize}
    \item \textbf{Data types:} we focus on integers in this work. But practical programs have booleans, strings, structs, classes, and so on. Variables assigned to values of different data types might be less likely to be swapped in WM.
    \item \textbf{Mental operations:} we focus on single-digit mental arithmetic with addition and subtraction. Harder mental operations may shift the balance e.g. in Experiment 2 and interfere with WM for program state.
    \item \textbf{Mutability}: we only test programs that assign to variables once, i.e. immutable state. Imperative programs often rely on in-place updates to databases, files, or so on. Mutating variables means updating an existing association between a variable/value in WM rather than adding a new one, which may be subject to different kinds of interference. Mutation also potentially increases the difficulty of on-demand tracing.
    \item \textbf{Control flow constructs:} we used functions due to their close relationship to on-demand tracing. Standard structured programming constructs like if-statements and for-loops should also be investigated for their WM influence.
    \item \textbf{Schematic structure:} our programs do not solve any common problem, e.g. sorting a list or computing the quadratic formula. A person's tracing strategies and WM encodings of program state would likely differ for programs with a recognizable structure, i.e. pieces that fit into a known schema.
\end{itemize}

\subsection{Implications for theory}

One goal of this work is to contribute towards a broader theory of program comprehension: how do people understand programs, and how do aspects of cognition influence that process? We focused on program tracing because it is a likely component skill of comprehension. We have shown that WM both limits how much state a person can remember, and causes different kinds of errors based on tracing strategies. However, most popular models of program comprehension focus largely on schemata/plans as a mechanism of comprehension, e.g. the seven models in the survey of von Mayrhauser and Vans\,\cite{von1995program}. We think that tracing deserves a closer look to understand its role in comprehension. 

One major finding of our work is that different people will adopt substantially different tracing strategies for the same program. We demonstrated that both the linear and on-demand strategies are common, and that individuals seem to prefer one strategy. But we do not yet know: what factors made a person pick one strategy over the other? And in what situations is a particular strategy better than the other? We found that linear tracing caused fewer WM errors than on-demand tracing for straight-line programs, but not for function programs. Future work should identify other factors that influence this trade-off.

Additionally, tracing as a task likely lies on a continuum of abstraction. We focused on tracing with concrete inputs and outputs, but one can imagine tracing over symbolic values, akin to the method of abstract interpretation in static analysis\,\cite{cousot1977abstract}. In this setting, variables are represented not by \textit{values} but by \textit{properties}. For example, if \verb|x > 0| and \verb|y = x + 1|, then we know \verb|y > 1|. Abstract tracing likely introduces even more WM load than concrete tracing. For example, while tracing ``\verb|if (x != NULL) { .. }|'', the property ``\verb|x != NULL|'' must be kept in working memory while reasoning about the body of the if-statement. A WM account of tracing should consider tracing at all levels of abstraction.


\subsection{Implications for design}

Another goal of our research is to generate design principles for reducing WM load that can be applied to IDEs, refactoring tools, style guides, and so on. We suggest a few guidelines that extend from our findings. As a general framing, it's important to observe that every experiment had significant between-subjects variability in memory capacity, time to completion, accuracy, strategy, and so on. Accessible programming tools need to account for working memory differences, e.g. tool designers should not assume that their users will be able to remember as much as themselves.

\subsubsection{Reduce variable scope.} If a program has many overlapping variables, it will be hard for a person to keep the variables' values in working memory as shown in Experiments 1 and 2. Therefore, programs should avoid having many variables within a given scope where possible. One implementation of this suggestion is to put a variable's definition as close as possible to its usage. This advice is consistent with most modern style guides, for example the Google C++ Style Guide\,\cite{googlestyle}. Our work confirms that this style has genuine cognitive benefits, as opposed to being a matter of preference.

This advice could also be mechanized as a complexity metric. The compiler technique of liveness analysis can identify when the live ranges (point from definition of variable to its last use) of variables will overlap. To test the technique, we applied it to every function in the Python standard library. We identified that some functions have up to 18 overlapping variables! We also found that functions with many parameters had a high complexity score, since they effectively are declaring dozens of variables at the top of a block rather than close to their usage, a de facto violation of the style rule.

\subsubsection{Visualize variable context.}

As shown in Experiment 3, people with different tracing strategies will make different kinds of WM errors, and so will need different kinds of tools to augment WM. A person tracing linearly needs to keep track of previously seen variables, and so a programming environment can visualize information about all the variables in scope up until a particular line. For example, the Lean interactive theorem prover\,\cite{de2015lean} in Emacs (Figure~\ref{fig:ides}, left) will show all the variables in scope at the current line based on the editor's mouse position. Lean only shows the name and type of the variable, but a visualization could also show the last line of code modifying that variable, or other semantic information.

By contrast, a person tracing on-demand needs to keep track of their path through the variable dependency graph so they can follow it back. If a person forgets a part of their path, a programming environment can visualize information about where a variable is used to help refresh their memory. For example, the DrRacket IDE\,\cite{findler2014drracket} (Figure~\ref{fig:ides}, right) will show all references to a variable on mouse hover. The point is that these ideas already exist in some form within (admittedly niche) programming tools, but \textit{every} IDE should support \textit{both} of these visualizations to reduce WM load during program tracing.




\subsubsection{Externalize program state.} 

Many of our experiments would be significantly different if the participants had been tracing on paper with a pen, able to write down program state such as variable values or flow markers. Providing programmers the ability to externalize their tracing process could significantly reduce working memory load.

However, nearly all interfaces for displaying code outside of an IDE are read-only. For example, GitHub has a specialized interface for pull requests that propose a change to a code base. The entire point of this interface is for a reviewer to understand a change, but the interface provides no ``margin notes'' or other features for throwaway annotations. Consistent with the theory of distributed cognition\,\cite{hutchins1995cognition}, code comprehension tools should consider how to enable programmers to externalize their thought process to the environment without needing to print out code.

\section{Conclusion}

In this work, we have shown that WM has a major role in maintaining program state during tracing. Beyond the basic predictions of WM theory (it is hard to remember a lot of state), we contribute a nuanced understanding of the type of WM errors a person can make. For example, programs involve associations between objects like variables and values, and WM errors involve not just forgetting but swapping these associations. A person will need to remember different kinds of information about a program based on their tracing strategy, and so they will make WM errors in different ways. With greater awareness of these issues, we can better design our programming tools to provide cognitive support for WM.

More broadly, we believe that our work demonstrates the enduring value of bridging the cognitive and computer sciences. Within HCI for programmers, cognitivist theory-building has largely fallen by the wayside in favor of contextual inquiry and system-building. But programming is a wicked problem, and cognitive science offers a unique lens for understanding challenges that involve (mostly) universal human capabilities. The last major cogsci contribution of HCI to programming, the Cognitive Dimensions of Notation\,\cite{green1989cognitive}, happened 30 years ago. It is high time to revitalize the goal of building a theory of how people program, one that spans from low-level cognitive resources to high-level sociological phenomena.

\bibliographystyle{ACM-Reference-Format}
\bibliography{sample-base}

\end{document}